\documentclass[twocolumn,aps,prb,superscriptaddress,showpacs,amsmath,amssymb,floats]{revtex4}

\usepackage{amsmath}
\usepackage{amssymb}
\usepackage{graphicx}
\usepackage{keyval}
\usepackage{dcolumn}
\usepackage{bm}
\usepackage{color}
\usepackage{txfonts}
\usepackage[]{sidecap}
\usepackage{epstopdf}
\usepackage{CJKutf8}
\usepackage{hyperref}

\begin{document}

\begin{CJK}{UTF8}{gbsn}


\title{The distinct in-plane resistivity anisotropy in the nematic states of detwinned NaFeAs and FeTe single crystals: evidences for Hund's rule metal}

\author{Juan Jiang (\CJKfamily{gbsn}姜娟), C. He(贺诚), Y. Zhang (张焱), M. Xu (徐敏), Q. Q. Ge (葛青亲), Z. R. Ye (叶子荣), F. Chen (陈飞), B. P. Xie (谢斌平)}
\author{D. L. Feng (封东来)}
\email{dlfeng@fudan.edu.cn}

\affiliation{State Key Laboratory of Surface Physics, Department
of Physics, and Advanced Materials Laboratory, Fudan University,
Shanghai 200433, People's Republic of China}

\date{\today}


\begin{abstract}

The  in-plane resistivity anisotropy has been studied with the Montgomery method on two detwinned parent compounds of the iron-based superconductors, NaFeAs and FeTe. For NaFeAs, the resistivity in the anti-ferromagnetic (AFM) direction is smaller than that in the ferromagnetic (FM) direction, similar to that observed in BaFe$_{2}$As$_2$ before. While for FeTe, the resistivity in the AFM direction is larger than that in the FM direction. We show that these two opposite resistivity anisotropy behaviors could be attributed to the strong Hund's rule coupling effects: while the iron pnictides are in the itinerant regime, where the Hund's rule coupling causes strong reconstruction and nematicity of the electronic structure; the FeTe is in the localized regime, where Hund's rule coupling makes hopping along the FM direction easier  than along the AFM direction, similar to the colossal magnetoresistance observed in some manganites.
\end{abstract}


\maketitle
\end{CJK}

\section{Introduction}

\begin{figure}[b!]
\includegraphics[width=7cm]{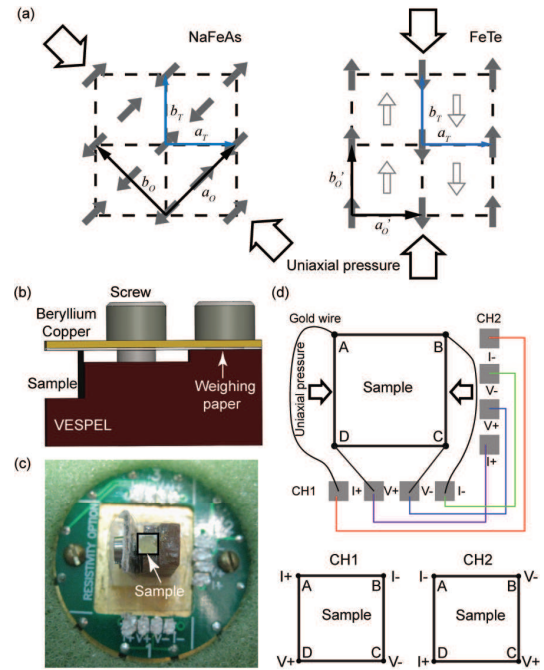}
\caption{(Color online) (a) The schematics of the spin structures in NaFeAs and FeTe following  Ref.~\onlinecite{ShiliangLi} and Ref.~\onlinecite{ShiliangLi2} , the  hollow arrows show the directions of the uniaxial pressure for detwinning. In NaFeAs, the $a_O$ ($b_O$) axis is 45 degree to the tetragonal $a_T$ ($b_T$) axis;\cite{ShiliangLi} while in FeTe, the $a_O\prime$ ($b_O\prime$) axis is the same as that in the tetragonal phase.\cite{ShiliangLi2} For both NaFeAs and FeTe, the spins are aligned antiferromagnetically along $a_O$/$a_O\prime$, and ferromagnetically along $b_O$/$b_O\prime$. (b) The design of the detwinning device, a piece of weighing paper is inserted between the sample and the Beryllium copper piece to ensure the insulation. (c) A detwinning device  mounted on a PPMS puck. (d) The two configurations in the Montgomery method for  in-plane resistivity measurement.} \label{Laue}
\end{figure}

Most unconventional superconductors are in the vicinity of certain magnetically ordered states. For cuprates,  the antiferromagnetic Mott insulator parental state is suggested to be intimately related to the superconducting mechanism. For iron-based high temperature superconductors (Fe-HTS),  several types of antiferromagnetic  parental states have been discovered, including the  collinear antiferromagnetic state (CAF) in iron pnictides,\cite{QHuang,PCDai,Pratt,Lester}  the bi-collinear antiferromagnetic state  (BCAF) in FeTe  [see Fig.~\ref{Laue}(a)],\cite{ShiliangLi2} the insulating block-antiferromagnetic state  of   K$_2$Fe$_4$Se$_5$, and a semiconducting collinear antiferromagnetic state in vacancy-ordered K$_x$Fe$_{1.5}$Se$_2$.\cite{Junzhao,ChenfeiPRX}

The collinear antiferromagnetic state breaks the four-fold symmetry, entering a nematic or two-fold symmetric phase, and it  was suggested to drive the tetragonal-to-orthorhombic structural transition as illustrated in Fig.~\ref{Laue}(a)  \cite{ChenFang,ChengHe}.   There are usually twinned domains in the orthorhombic states, but it has been shown that the twinning could be removed with a uniaxial pressure.  In  detwinned BaFe$_{2-x}$Co$_x$As$_2$, the resistivity in the AFM  ($a_O$)  direction is found to be significantly smaller than that in the FM  ($b_O$) direction.\cite{JHChu}  Such a resistivity anisotropy could be taken as a hallmark of  the nematic phase.  Later on, the resistivity anisotropy was shown to be much reduced for the post-annealed BaFe$_{2-x}$Co$_x$As$_2$.\cite{TLiang} On the other hand, the resistivity anisotropy was found to be much weaker in detwinned Ba$_{1-x}$K$_x$Fe$_2$As$_2$.\cite{Ying} Theoretically,
some suggest that the resistivity anisotropy is an indication for the presence of orbital ordering,\cite{Kuorbitalorder}  while others  suggest that the details of the quasi-particle scattering and Fermi surface topology might be responsible for the  diversified behaviors in the electron and hole doped BaFe$_{2}$As$_2$.\cite{Rafaelxx}

\begin{SCfigure*}[][t!]
\includegraphics[width=11cm]{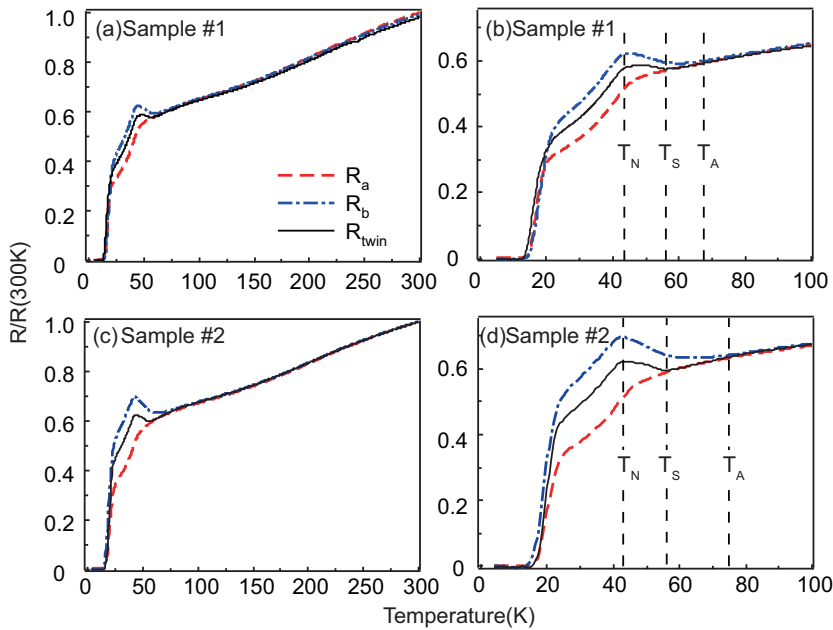}
\caption{(Color online) (a) Temperature dependence of the normalized in-plane resistance of NaFeAs single crystal measured with   the Montgomery method . $R_a$ (dashed curve) and $R_b$ (dash-dotted curve) are the in-plane resistances along the $a_O$ and $b_O$ directions respectively  when the uniaxial pressure is applied along the $b_O$ direction for detwinning.  $R_{twin}$ (thin solid curve) is measured on the same sample when the pressure is not applied, thus it corresponds to the resistivity of the twinned crystal.  (b) is an enlargement of panel (a). (c) and (d) are the same as panels (a) and (b) respectively, but for a different sample. }\label{NaFeAs}
\end{SCfigure*}

Previous studies were limited to the so-called ``122" series of Fe-HTS, we here report the  temperature dependence of resistivity anisotropy in the ``111" and ``11" series of Fe-HTS, specifically, in NaFeAs and FeTe single crystals detwinned with uniaxial strain. For NaFeAs, we found that the resistivity in the AFM direction is smaller than that in the FM direction, similar to the BaFe$_{2-x}$Co$_x$As$_2$ case.\cite{JHChu}  Intriguingly, for FeTe, the resistivity anisotropy exhibits an opposite behavior to that of NaFeAs. That is, the resistivity along the AFM direction is larger than that along the FM direction. Considering the electronic structures measured by angle-resolved photoemission spectroscopy (ARPES), we propose that the dramatically difference in the resistivity anisotropy behaviors of NaFeAs and FeTe are caused by the different manifestation of the  Hund's rule coupling ($J_H$) in the itinerant regime and localized  regime, respectively.
For NaFeAs, the anisotropy is caused by nematicity of the electronic structure reconstruction induced by $J_H$,  while for FeTe, carriers are localized, and $J_H$ induces extra potential barrier for the electrons to hop in the AFM direction, and thus higher resistance than in the FM direction. This is analogous to the hopping facilitated by the ferromagnetic spin orientation in the colossal  magnetoresistance (CMR) effect of certain manganites.\cite{CMR} Our results suggest that while on-site Coulomb interaction U may not be strong for Fe-HTS, $J_H$ is the dominating interaction that gives strong on-site correlations in such a multi-orbital system, just like in the manganites.

\begin{figure*}[t!]
\includegraphics[width=16cm]{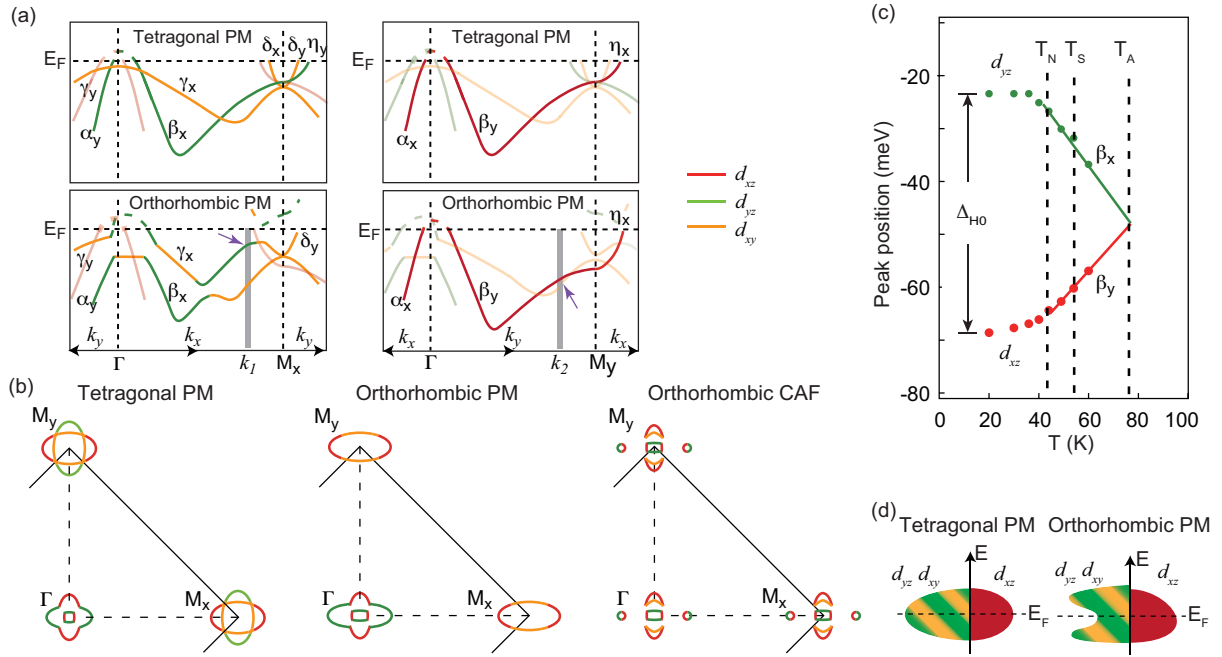}
\caption{(Color online) The electronic structure extracted from ARPES measurements of NaFeAs. (a) The band structure in  the tetragonal paramagnetic state, and  the orthorhombic paramagnetic state (nematic state) along the $k_x$ (or $a_{O}$) and $k_y$ (or $b_{O}$) directions.  Note the bands are labelled as if there were no hybridizations.  For simplicity, some bands are plotted with weak intensities.  The maximal separation between the $\beta_x$ and $\beta_y$ bands are found at the  momenta $k_1$ and $k_2$ respectively, as marked by the arrows. (b)  The Fermi surface evolution in  the tetragonal paramagnetic, orthorhombic paramagnetic, and orthorhombic collinear antiferromagnetic states. (c) The maximal separation between the $\beta_x$ and $\beta_y$ bands as a function of temperature. (d) The artistic sketch of the occupation alterations of various orbitals in the tetragonal paramagnetic and orthorhombic paramagnetic states. }\label{arpes}
\end{figure*}

\section{Experimental}

The NaFeAs and $\alpha$-FeTe  single crystals were synthesized following previous reports.\cite{ChengHe,Taen} There are $\sim$8.8~\% excess Fe in FeTe based on our energy-dispersive x-ray spectroscopy (EDX) data. Since the as-grown crystals contain twin domains in the orthorhombic phase, we designed a similar detwinning device as that of Chu \emph{et al.} [see Fig.~\ref{Laue}(b)],\cite{JHChu} which puts a uniaxial pressure along one of the orthorhombic direction. It has been shown that the twinning in iron pnictides could be removed  effectively in this way,\cite{fisherreview} and   $b_O$   is preferred along the pressurized direction in the orthorhombic phase, since the lattice constant $b_O$ is slightly smaller than $a_O$.\cite{Tanatardomain} As shown in Fig.~\ref{Laue}(c),  a single crystal was cut to a rectangular shape with edges along the $a_O$ and $b_O$ axes determined through its  Laue x-ray diffraction pattern, and mounted on the detwinning device.  We found that this device could effectively remove twinning in both NaFeAs and FeTe, since  $b_O^\prime$ is smaller $a_O^\prime$ in FeTe as well.
The resistivity measurements were conducted with a Quantum Design physical property measurement system (PPMS), using the Montgomery method as shown in Fig.~\ref{Laue}(d).\cite{Montgomery,Logan,TLiang} This method has the advantage to obtain  the resistances along both orthorhombic directions at the same condition, compared with the usual four-lead method.\cite{TanatarBaCa,Blomberg}

\section{the resistivity anisotropy of NaFeAs}

The electric resistance of the twinned NaFeAs, $R_{twin}$,  is measured by the  Montgomery method, before the uniaxial pressure is  applied. As shown in Figs.~\ref{NaFeAs}(a) and ~\ref{NaFeAs}(b), it is metallic at high temperatures, and exhibits an upturn at the structural transition temperature $T_S$ ($\sim$56~K).\cite{GFChen} There is a  hump around the CAF transition temperature $T_N$  ($\sim$43~K), before it drops to zero rapidly, due to the filamentary superconductivity induced by a slight Na deficiency. Consistently,  magnetic susceptibility shows a negligible superconducting volume fraction for these NaFeAs samples (data not shown here).

When a uniaxial pressure is applied along $b_O$ as shown in Fig.~\ref{Laue}(a), the resistance along the $a_O$ direction, $R_a$, remains metallic, but $R_b$ along the $b_O$ direction starts to increase at the temperature $T_A$,  about 12~K above $T_S$,   exhibiting an anisotropy with $R_b>R_a$ [see Fig.~\ref{NaFeAs} (b)]. Similar behaviors of the in-plane resistivity anisotropy were observed in the ``122" series of iron pnictides.\cite{JHChu,TanatarBaCa,Blomberg} It indicates that such resistivity anisotropy  in the nematic state is not a unique  feature of the ``122" series of iron pnictides. We note that the resistance drop due to filamentary superconductivity in $R_a$ or $R_b$ is at  a slightly higher temperature than that in $R_{twin}$, and it is likely due to the creation of  additional filamentary superconducting routes by pressure  in this particular sample .

In Figs.~\ref{NaFeAs}(c) and ~\ref{NaFeAs}(d), we present data taken on another sample from the same batch. The  anisotropy is larger than that of the first sample, and $R_b$ starts to deviate from $R_{twin}$ and $R_a$ at a higher $T_A$ of 75~K, which  suggests that the strain and thus the degree of detwinning  should be larger for the second sample.    However, $R_a$ still starts to deviate from  the strain-free $R_{twin}$   at the same $T_S$ as the first sample, and the CAF transition temperature $T_N$ seems to be unaffected as well.  It is still to be understood why $R_b$  is more sensitive to the strain than $R_a$.

The onset of the resistivity anisotropy, or nematicity, is clearly above the structural and CAF transitions in NaFeAs, which is similar to that observed in BaFe$_{2-x}$Co$_x$As$_2$.
Recently, nematicity has been demonstrated by magnetic torque well above $T_S$ in an unstrained microscopic BaFe$_2$(As$_{1-x}$P$_x$)$_2$ sample with unbalanced twin domains.\cite{Matsudanematicity} An early ARPES study on an unstrained NaFeAs has found that there is a reconstruction of the electronic structure that is directly related to the nematic transition, and it occurs at a temperature slightly above $T_S$.\cite{ChengHe} Therefore, the nematicity above $T_S$ is thus an intrinsic property. Our findings indicate that the nematicity can be further enhanced by the uniaxial strain. Consistently, a recent neutron scattering experiment demonstrates that the collinear magnetic fluctuations emerge  at a higher temperature  in the mechanically detwinned, \textit{i.e.} strained, BaFe$_2$As$_2$ than in the strain-free sample. Moreover,  the structural transition is smeared to higher temperatures by mechanically detwinning,\cite{Blomberg2} and the resistivity anisotropy exhibits a long ``tail" above $T_N$/$T_S$.

\begin{figure}[t!]
\includegraphics[width=6cm]{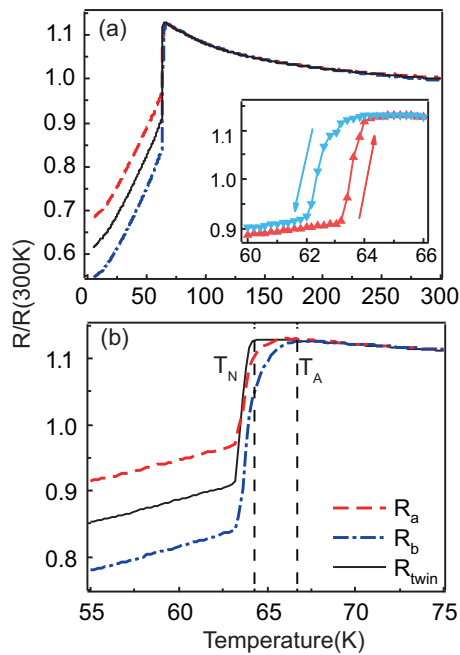}
\caption{(Color online) (a) Temperature dependence of the normalized in-plane resistance of FeTe single crystal. $R_{twin}$ (thin solid curve) is the resistance of the twinned sample before any pressure is applied, while $R_a$ (dashed curve) and $R_b$ (dash-dotted curve) are the in-plane resistance along the $a_O\prime$ and $b_O\prime$ directions after the uniaxial pressure is applied for detwinning.  (b) is the enlargement of panel (a).   All the data were measured during warming the sample, except those in the inset of panel (a), where  a  hysteresis loop of the resistance is observed  with a width of about 1 K.
 }\label{FeTe}
\end{figure}

\begin{figure*}[t!]
\includegraphics[width=12.6cm]{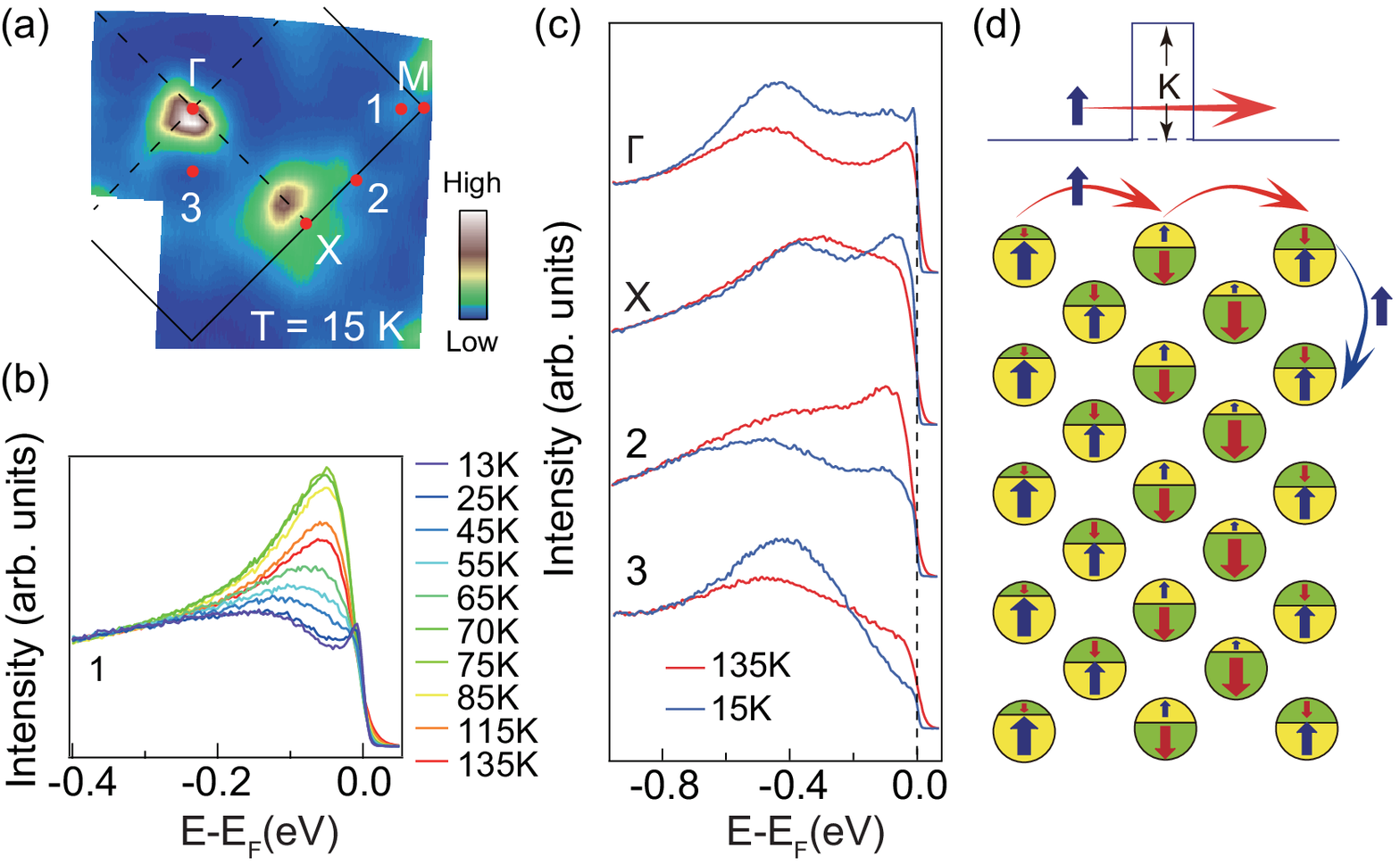}
\caption{(Color online)   (a) Photoemission intensity
distribution integrated over the energy window of $[E_F-15~meV, E_F+15~meV]$ for Fe$_{1.06}$Te measured at 15~K.  (b) Detailed temperature dependence of photoemission spectrum at the momentum position \#1 as marked in panel (a).  (c)
Temperature dependence of photoemission spectra at various momenta as marked in panel (a).
(d) A schematic local picture of the bi-collinear antiferromagnetic order in FeTe, where the arrows represent  spins, the size of the arrows and the area that they occupy represent their population and size of the moment.  Two hopping routes are demonstrated. When an up-spin electron hops along the AFM direction, it feels an additional potential $K$ on the site with the localized moment pointing down.  On the other hand, it moves freely in the FM direction. }\label{FeTe_arpes}
\end{figure*}

Various mechanisms have been proposed for the observed in-plane resistivity anisotropy, including the structural orthorhombicity, the collinear spin order.\cite{QHuang,JunZhao,ShiliangLi}. and the nematicity in the electronic structure (either the Fermi surface topology\cite{Rafaelxx}, or the orbital ordering \cite{Kuorbitalorder}). In the following, they will be discussed one by one. \\
\textit{i) Structural orthorhombicity.}  The relative difference between $a_O$ and $b_O$ is only $\sim$0.3\%  in NaFeAs, which is less likely the cause of the large anisotropy of $\sim$30\%.  \\  \textit{ii) Collinear spin order.}  This pure spin scenario requires  significant difference  in the quasi-particle scattering rates along the FM and AFM directions. However, it has not been observed.\cite{NaFeAs_yzhang}\\
 \textit{iii) Electronic structure nematicity.} Recently, our polarization-dependent ARPES study on a uniaxially strained NaFeAs has clearly demonstrated the orbital dependent  electronic structure reconstruction  in the nematic state.\cite{NaFeAs_yzhang} In Fig.~\ref{arpes}(a), we have reproduced the measured electronic structure in the tetragonal paramagnetic state above $T_A$, and  that in the orthorhombic paramagnetic (nematic) state between $T_A$ and $T_N$ along both the a$_O$ and b$_O$ directions.  In the CAF  (also nematic) state, additional folding  would complicate the electronic structure, as shown for the Fermi surface in Fig.~\ref{arpes}(b), however, the main dispersions are similar to those in the orthorhombic paramagnetic state and thus not shown here.\cite{ChengHe} The electronic structure reconstructs dramatically across the phase transitions. Particularly, the   $\beta$ band exhibits the most remarkable reconstruction. With decreased temperature, the $\beta$ band dispersion along the AFM direction, or $\beta_x$, is pushed up toward $E_F$,  while the dispersion along the FM direction ($\beta_y$) is pushed  to higher binding energies [Fig.~\ref{arpes}(a)]. The population of $\beta_x$ is reduced, when it crosses $E_F$, thus the total energy could be reduced.   Because such a shift happens on the entire band, it gives a much larger energy gain than the opening of the  hybridization gap near $M_x$.
  The electronic structure reconstruction  is thus suggested to drive the phase transitions.\cite{ChengHe}
Since the band structure and Fermi surface reconstruct dramatically, and exhibit an  anisotropic behavior as shown in Fig.~\ref{arpes}(b), it would be sufficient to account for the resistivity anisotropy.   Consistently, as shown in Fig.~\ref{arpes}(c), the electronic structure reconstruction  could be traced to  75~K, exactly the same as the $T_A$ in the sufficiently detwinned NaFeAs sample \#2.\cite{NaFeAs_yzhang,NaFeAs_MYi,MingYi}  On the other hand,  orbital ordering has been proposed to explain the resistivity anisotropy,\cite{Kuorbitalorder,Shinorbitalorder} where the occupation of the $d_{xz}$ orbital that extends in the AFM direction significantly increases and enhances hopping along that direction, while occupation of the $d_{yz}$ orbital decreases. However, we found that the orbital occupations in the high temperature tetragonal phase and low temperature nematic phase only differ by less than 5\% [see Fig.~\ref{arpes}(d)], which is much less than what the orbital ordering scenario expects.\cite{Kuorbitalorder,Shinorbitalorder} Thus the orbital ordering is unlikely the origin for such a large resistivity anisotropy.  \\
Therefore, based on above arguments, one could conclude that  the resistivity anisotropy are most likely induced by the observed nematic band structure and Fermi surface, and  the exact nature of the anisotropy might depend on the details in the anisotropic quasiparticle scattering rate combined with the Fermi surface topology.

The remarkable electronic structure reconstruction was proposed to be conspired by both the Hund's rule coupling and the CAF order or CAF fluctuations at high temperatures,\cite{LXYang,NaFeAs_yzhang} which is observed in the neutron scattering experiment.\cite{Blomberg2} The finite local moments interact with an itinerant band through the Hund's rule coupling $J_H$, and shift  the energy of the parallel and anti-parallel spins differently. Note that since the spin up and spin down sites are equally present, the bands are still populated by electrons with both spins.   In the recent dynamic mean field theory (DMFT) studies, such electronic structure reconstruction is reproduced when Hund's rule coupling are included.\cite{Yin1,Yin2} Therefore, the observed anisotropic resistivity of NaFeAs is essentially originated from the Hund's rule couplings in this system.

To give a characteristic energy scale of such a Hund's rule coupling effect,  we define the maximal observable separation between $\beta_x$ and $\beta_y$ at the same momentum value (\textit{i.e.} $|k_x|=|k_y|$) near $M_x$ and $M_y$ respectively as $\Delta_{H}$.  $\Delta_{H}$ is a function of temperature, and its low temperature saturated value is defined as $\Delta_{H0}$  [see Fig.~\ref{arpes}(c)].

\section{the resistivity anisotropy of  FeTe}

The resistance of twinned FeTe is shown in Fig.~\ref{FeTe}(a) as a function of temperature. It exhibits a semiconducting behavior in the paramagnetic state, and then becomes metallic after a sharp drop around 64~K, corresponding to the structural/BCAF transition.\cite{MHFang} Moreover, the hysteresis loop is observed in the resistance [see the inset of Fig.~\ref{FeTe}(a)] with the width about 1~K, revealing its first-order nature.\cite{GFChen2,ShiliangLi2} When a uniaxial pressure is applied along the $b_O\prime$ direction  of a FeTe single crystal with the detwinning device, both $R_a$ and $R_b$ still exhibit sharp drops around $T_N$, however, the resistivity anisotropy appears below $T_A$, which is  about 5~K above $T_N$ as shown in Fig.~\ref{FeTe}(b).   Most strikingly, we found that the anisotropy in FeTe exhibits an opposite behavior to that in NaFeAs, that is, $R_a$ in the AFM direction is larger than $R_b$ in the FM direction for FeTe.

Our early ARPES measurements\cite{YZhang}  have illustrated the polaronic nature of  FeTe.  As shown in Fig.~\ref{FeTe_arpes}(a) for a twinned Fe$_{1.06}$Te sample, the photoemission intensity around $E_F$ is distributed as large patches over a broad momentum region, and the  Fermi surface is poorly defined. In the high temperature paramagnetic state, the photoemission spectrum is broad, and the single particle excitation spectral function is overwhelmed by incoherent spectral weight  [Fig.~\ref{FeTe_arpes}(b)], while a small but sharp quasiparticle peak emerges in the BCAF state.  The incoherent spectral weight  is responsible for the  semiconducting behavior of the  resistivity at high temperatures, and the small but coherent quasiparticle gives the metallic behavior below $T_N$. Fig.~\ref{FeTe_arpes}(c) compares several typical spectra in the paramagnetic state and BCAF state at various momenta in the Brillouin zone, which illustrate that the incoherent spectral weight is relocated over a large momentum and energy phase space across the BCAF transition. In general, it was found that the spectral weight is suppressed in the [$E_F$-0.4~eV, $E_F$] region, and enhanced in the [$E_F$-0.7~eV, $E_F$-0.2~eV] region, which would significantly save the electronic energy, and thus is sufficient to drive the phase transition.  \cite{YZhang}  These  distinct electronic properties  show that FeTe possesses the most localized and polaronic characters among all the iron-based compounds; and consistently, it has the large moment of $\sim$2 $\mu_B$ per Fe site as observed in  neutron scattering experiments.\cite{PCDai2}  Therefore,   FeTe could be better understood from a local picture as sketched in Fig.~\ref{FeTe_arpes}(d) and elaborated in the caption.\cite{WeiGuoYin}  In this picture, the Hund's rule coupling between the itinerant electron (the small quasiparticle weight in the single particle excitation spectrum, with spin 1/2) and the localized moment ($S$) would enforce an additional barrier $K$ (proportional to $J_H S$) for the electrons hoping along the AFM direction, whereas  those hopping along the FM direction are free of such a barrier [see Fig.~\ref{FeTe_arpes}(d)].  This naturally explains the observed resistivity anisotropy of FeTe.


Similar physics (multi-band, polaronic electronic structure with strong Hund's rule coupling) has been observed before in the manganites \cite{Mannella}. Actually, a similar temperature dependence  of resistivity is observed in LaMnO$_3$ doped with bivalent cations: above the Curie temperature, the resistivity behaves like a semiconductor, then a metallic conduction is observed in the ferromagnetic phase.\cite{Urushibara} It has been well comprehended by the double-exchange  model, where  electrons hop freely when the local moments are ordered ferromagnetically. In the paramagnetic state,  a random potential is enforced by  $J_H$ at sites whose moments are not parallel to the spin of the hopping electrons. Such a randomness would cause localization and thus the insulating behavior.  Therefore, just like the colossal magnetoresistivity,   the resistivity anisotropy of FeTe is a remarkable evidence of strong appearance of Hund's rule coupling.

\section{Discussion and conclusion}

\begingroup
\begin{figure}[t!]
\includegraphics[width=8.6cm]{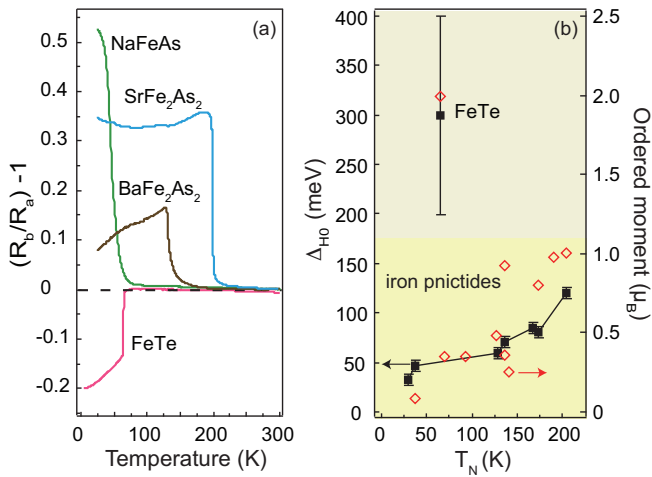}
\caption{(Color online) (a) Temperature dependencies of $R_b/R_a-1$ in BaFe$_2$As$_2$, SrFe$_2$As$_2$, NaFeAs, and FeTe. The resistivity anisotropy of BaFe$_2$As$_2$ was also measured with the Montgomery method. The data for SrFe$_2$As$_2$ were taken from Ref.\onlinecite{Blomberg} . (b)  $\Delta_{H0}$ obtained from our ARPES data (including both the published and the unpublished), and the low temperature ordered moment measured by neutron scattering in various iron pnictides are plotted as a function of the Neel temperature, and the data are also tabularized in the following:
}\label{sum}
\end{figure}

 \begin{table}
\begin{tabular}{c|ccc}
Compound & $T_N$(K) & \hspace{10pt} $\Delta_{H0}$ (meV) \hspace{10pt}  & \hspace{10pt} $S$   ($\mu_B$) \hspace{10pt} \\    \hline
\textbf{FeTe} & \textbf{67} & \textbf{300} & \textbf{2.0}  (Ref.~\onlinecite{ShiliangLi2})  \\
SrFe$_2$As$_2$ & 205 & 120 (Ref.~\onlinecite{YZhang2}) & 1.01(3) (Ref.~\onlinecite{SrFeAs})  \\
EuFe$_2$As$_2$ & 190 & - & 0.98(8) (Ref.~\onlinecite{EuFeAs})  \\
CaFe$_2$As$_2$ & 173 & 80 & 0.80(5) (Ref.~\onlinecite{CaFeAs}) \\
Sr$_{0.9}$K$_{0.1}$Fe$_2$As$_2$ & 168 & 85  (Ref.~\onlinecite{YZhang2})& - \\
NdOFeAs & 141 & - & 0.25(7) (Ref.~\onlinecite{NdOFeAs})  \\
LaOFeAs & 137 & - & 0.36  (Ref.~\onlinecite{PCDai})  \\
BaFe$_2$As$_2$ & 136 & 70  (Ref.~\onlinecite{MYi}) & 0.93(6) (Ref.~\onlinecite{BaFeAs})  \\
Sr$_{0.82}$K$_{0.18}$Fe$_2$As$_2$ & 129 & 60  (Ref.~\onlinecite{YZhang2}) & -  \\
PrOFeAs & 127 & - & 0.48(9) (Ref.~\onlinecite{PrOFeAs})  \\
BaFe$_{0.95}$Co$_{0.05}$As$_2$ & 93 & - & 0.35  (Ref.~\onlinecite{BaCo})  \\
Ba$_{1-x}$K$_x$Fe$_2$As$_2$ & 70 & - & 0.35  (Ref.~\onlinecite{BaK})  \\
NaFeAs & 37 & 46  (Ref.~\onlinecite{NaFeAs_yzhang}) & 0.09(4) (Ref.~\onlinecite{NaFeAs})  \\
NaFe$_{0.9825}$Co$_{0.0175}$As & 30 & 32  (Ref.~\onlinecite{NaFeCoAs}) & -
\end{tabular}
 \end{table}
\endgroup

The resistivity anisotropy properties  for NaFeAs, BaFe$_2$As$_2$, SrFe$_2$As$_2$ and FeTe, presented through $R_b/R_a-1$, are summarized in Fig.~\ref{sum}(a).  As expected, the iron pnictides  exhibit a similar behavior,  since  they share the same CAF ground state, similar normal-state electronic structures, and common band reconstructions in their CAF states.\cite{ChengHe,YZhang2,LXYang,MingYi} Moreover, FeTe and SrFe$_2$As$_2$ exhibit sharp resistivity anisotropy jump at $T_N$, reflecting the strong first order nature of the phase transition. Relatively,  $T_A$ does not extend too far above $T_N$ for both of them,  compared the long tails in the resistivity anisotropy  of  NaFeAs and BaFe$_2$As$_2$.\cite{TanatarBaCa,Blomberg} It suggests that the nematic fluctuation is suppressed for the first order phase transition case, whereas it is rather strong above the weak first order or second order phase transition temperatures in NaFeAs and BaFe$_2$As$_2$.

From the electronic structure perspective, the CAF/BCAF phase transitions are driven by the energy gain during the band structure reconstruction in iron pnictides or large scale spectral weight transfer in FeTe. These two different forms of electronic structure construction are  the direct consequences of the Hund's rule coupling in different regimes. Namely,  iron pnictides is in the itinerant regime, and FeTe is in the  localized regime.  One  could distinguish $S$ as ``local moment" in iron pnictides, and as  ``localized moment" in FeTe. Local moment in iron pnictides is the net moment due to the population difference between the majority and minority bands projected  onto a particular site, which can be obtained by integrate the spin-dependent Bloch wavefunctions  around certain  site in the itinerant picture. For FeTe, most of the electrons can be considered localized, and there is just a very small coherent quasiparticle weight. Theoretically, both the localized regime and the itinerant regime were unified by a recent DMFT+DFT (density functional theory) calculation that considers the Hund's rule coupling as the most important local correlations.\cite{Yin1,Yin2} The band structure reconstruction was reproduced, although it was not exactly like the experiment; and the saving of electronic energy through $J_H$ is found to be the dominating force behind the nematic phase transitions.  Moreover, the $J_H$ effects are  found outstanding for FeTe amongst various iron-based compounds. On a similar footing,  a model containing both the itinerant electrons and local moments was proposed  to unify the magnetic ground states in both the iron pnictides and FeTe.\cite{WeiGuoYin} Besides the various exchange interactions amongst neighboring sites,\cite{Luzhongyi,WeiBao}it was proposed that the decisive parameter for magnetic order is  the energy barrier $K$ for the electron hopping between the antiparallel sites as depicted in Fig.~\ref{FeTe_arpes}(d), which is determined by  the Hund's rule coupling energy. Both the CAF order in iron pnictides and the BCAF order in FeTe could be present in the same phase diagram and theoretical framework.\cite{WeiGuoYin} Based on these theories,  and our electronic structure and resistivity anisotropy measurements, both NaFeAs  and FeTe could be considered as the Hund's  metals.

To put other iron pnictide parental compounds into the Hund's metal picture discussed here,  we further examine the general relation between  representative electronic and magnetic quantities and the CAF/BCAF Neel temperatures. We take
the low-temperature saturated value of the $\beta$ band reconstruction energy scale, $\Delta_{H0}$, as a representative of the  energy scale of  CAF transition, which should scale with  $J_H S$.   Fig.~\ref{sum}(b) collects $\Delta_{H0}$ for various iron-based compounds as a function of $T_N$. They indeed show certain monotonic correlations.
On the other hand,  the band reconstruction in FeTe appear as the spectral weight transfer over a broad energy range  in the entire Brillouin zone  [see Fig.~\ref{FeTe_arpes}], due to its  polaronic nature. One  could roughly estimate $\Delta_{H0} \sim 0.3\pm0.1$~eV for FeTe,  much larger than  $\Delta_{H0}$'s of  iron pnictides with similar $T_N$.
 Overlaid on the same Fig.~\ref{sum}(b) are the ordered moments measured by neutron scattering experiments at  low temperatures as a function of $T_{N}$.\cite{SrFeAs, EuFeAs, CaFeAs, PCDai, NdOFeAs, BaFeAs, PrOFeAs, BaCo, BaK, ShiliangLi2, NaFeAs} Although there are sizable variations,  it is intriguing that the  ordered moment $S$ generally follows $\Delta_{H0}$, indicating that $J_H$ does not vary too much amongst various iron-based compounds. Intriguingly, the moment of FeTe  ($\sim 2\mu_B$) is again out of the scale, compared with  those of the iron pnictides. The ordered moment  could be taken as the order parameter of the antiferromagnetic phase transition, which arises below $T_N$.  However, the band reconstruction starts at $T_A$ (above $T_N$). This is because the time scale of ARPES is very fast ($\sim 1$~fs), which could capture the fluctuating and short-ranged nematicity, while quasi-elastic neutron scattering gives a time-averaged ordered moment over a time scale much longer than 1~ps.  Therefore,  they reflect the different aspects of the same physics. Particularly,  $\Delta_H$ might be taken as an order parameter for the electronic nematicity, which was shown to be a true phase transition that occurs at $T_A$ by the magnetic torque experiments for
BaFe$_2$(As$_{1-x}$P$_x$)$_2$. \cite{Matsudanematicity}
 The remarkable correlation between the ordered moment, $\Delta_{H0}$, and $T_N$ for various iron pnictides and FeTe indicate the general applicability of the Hund's metal physics for the iron-based compounds.

To conclude, we report the temperature-dependence of the resistivity anisotropy in detwinned NaFeAs and FeTe single crystal by Montgomery method. The resistivity anisotropy of NaFeAs is intrinsic and resembles that of the ``122" series of iron pnictides, while FeTe shows an opposite behavior. We find that they both could be understood by considering the different manifestations of the Hund's rule coupling in the itinerant and localized limit respectively. Moreover, we present a  plot between $\Delta_{H0}$, the ordered moment $S$, and the Neel temperature $T_N$ of various iron-based compounds in the antiferromagnetic state, which highlights the critical role of the Hund's rule coupling in the electronic nematicity, magnetic order, and in the generic physics of Fe-HTS.

\textit{Acknowledgement:} We acknowledge the helpful discussions issues with Dr. Wei Ku, Mr. T. Liang and Mr. S. Ishida. This work is supported in part by the National Science Foundation of China, Ministry of Education of China, and National Basic Research Program of China (973 Program)  under the grant Nos. 2011CB921802, 2011CBA00112, and 2012CB921400.

\end{document}